\documentclass{article}
\usepackage{ijcai24}

\usepackage{times}
\usepackage{soul}
\usepackage{url}
\usepackage[hidelinks]{hyperref}
\usepackage[utf8]{inputenc}
\usepackage[small]{caption}
\usepackage{graphicx}
\usepackage{amsmath}
\usepackage{amsthm}
\usepackage{booktabs}
\usepackage{algorithm}
\usepackage[switch]{lineno}

\usepackage{amssymb}  
\usepackage{algpseudocode}

\usepackage{xcolor}
\colorlet{RED}{red}

\title{Closed-loop Teaching via Demonstrations to Improve Policy Transparency}

\author{
    Michael S. Lee, Reid Simmons, Henny Admoni
    \affiliations
    Robotics Institute, Carnegie Mellon University
    \emails
    \{ml5, rsimmons, hadmoni\}@andrew.cmu.edu
}

\begin{document}

\maketitle

\begin{abstract}
Demonstrations are a powerful way of increasing the transparency of AI policies. Though informative demonstrations may be selected \textit{a priori} through the machine teaching paradigm, student learning may deviate from the preselected curriculum \textit{in situ}. This paper thus explores augmenting a curriculum with a closed-loop teaching framework inspired by principles from the education literature, such as the zone of proximal development and the testing effect. We utilize tests accordingly to close to the loop and maintain a novel particle filter model of human beliefs throughout the learning process, allowing us to provide demonstrations that are targeted to the human's current understanding in real time. A user study finds that our proposed closed-loop teaching framework reduces the regret in human test responses by 43\% over a baseline.
\end{abstract}

\section{Introduction}
\label{sec:introduction}

Much progress has been made in obtaining complex and capable AI policies through reinforcement learning (e.g. \cite{brohan2023can}). Ensuring the transparency (i.e. understandability and predictability \cite{endsley2017here}) of these policies in all scenarios is key to calibrating the expectations of developers and end-users toward proper usage; however, this remains a challenge \cite{wells2021explainable}.

One effective way to increase policy transparency is through demonstrations of the policy, which can be selected through a \textit{machine teaching} \cite{zhu2015machine} paradigm that selects the minimal set of examples (e.g. demonstrations) that will help a student comprehend a concept (e.g. a policy) given their learning model. Though machine teaching can assist in selecting a principled curriculum of demonstrations \textit{a priori}, student learning may deviate from the modeled learning trajectory \textit{in situ}. 
In prior work by Lee et al.~\shortcite{lee2022reasoning}, machine teaching-selected demonstrations improved human performance on post hoc tests examining understanding of later-demonstrated concepts but decreased performance on post hoc tests examining understanding of early-demonstrated concepts, suggesting perhaps that the curriculum moved too quickly past the early concepts without in situ testing to provide additional instruction as necessary.

\begin{figure}[t]
\centering
\includegraphics[width=0.94\columnwidth]{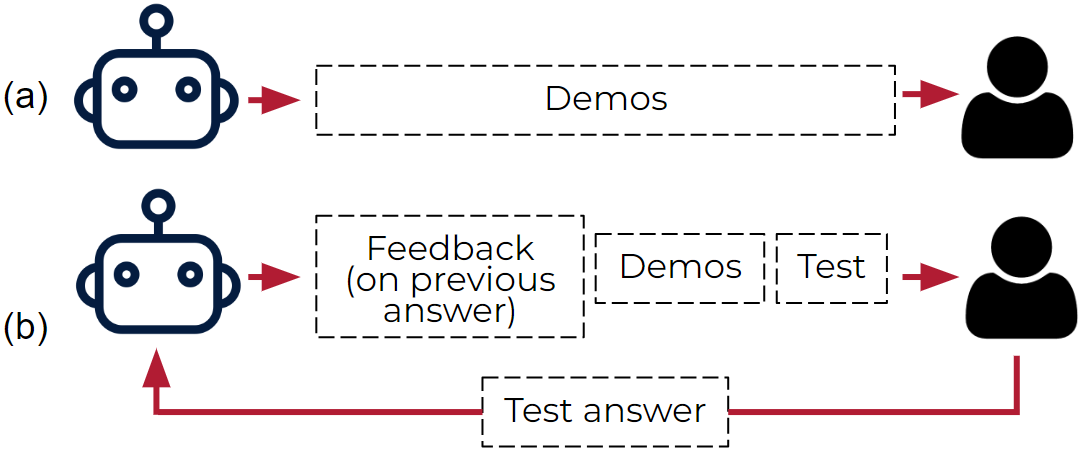} 
\caption{(a) Previous works aim to improve policy transparency via a set of demonstrations selected \textit{a priori}, but student learning may deviate from the expected trajectory. (b) We propose a closed-loop teaching framework using tests and feedback to detect and correct for such deviations \textit{in situ}.}
\vspace{-5mm}
\label{fig:overview}
\end{figure}

Thus, \textit{our key idea is to complement a curriculum of machine teaching-selected demonstrations with a closed-loop teaching framework inspired by the education literature to provide tailored instruction in real time} (Fig \ref{fig:overview}). A guiding educational concept is teaching in the \textit{zone of proximal development} (ZPD) or ``Goldilocks zone'' \cite{hattie2018visible,vygotsky1980mind}, which suggests that the examples provided to the learner should not be too easy nor too difficult, given their current understanding. However, the ZPD often changes at different rates for different students based on their personal learning rate, which must be assessed periodically through testing. We inform the testing cadence with the educational concept of the \textit{testing effect} \cite{roediger2006power}, which predicts an increase in learning outcomes when a portion of the teaching budget is devoted to testing the student (leveraging testing not only as a tool for assessment but also for teaching). And by incorporating tests and feedback in a closed teaching loop, we maintain an up-to-date model of human beliefs and promote demonstrations that are provided at the right level of difficulty in situ.


\begin{figure*}[!t]
\centering
\includegraphics[width=0.85\textwidth]{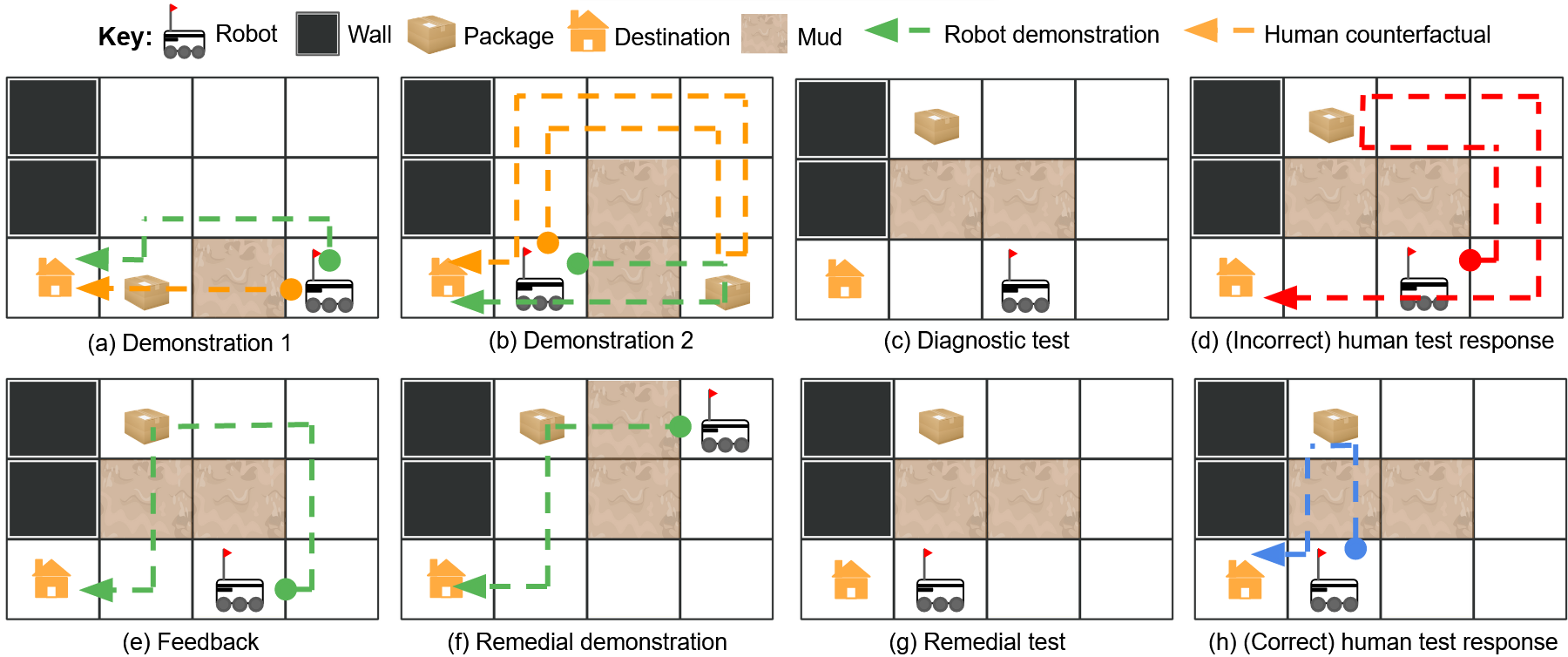} 
\caption{Sample teaching sequence for a batch of KCs on mud cost. \textbf{(a)} First demonstration (green) contrasts with a counterfactual alternative likely considered by a human (orange), which conveys that mud is costly. \textbf{(b)} Second demonstration lowerbounds mud cost. \textbf{(c)} Human is asked to predict the robot's behavior in a test. \textbf{(d)} Incorrect response suggests that the demonstration was not understood. \textbf{(e)} Human is given the correct response as feedback. \textbf{(f)} Remedial demonstration is provided to target the misunderstanding. \textbf{(g)} Human is given a remedial test. \textbf{(h)} Correct answer suggests understanding.}
\label{fig:combined}
\vspace{-5mm}
\end{figure*}

To illustrate the utility of our closed-loop teaching framework, consider a robot that increases the transparency of its reward function and policy to a human using demonstrations, tests, and feedback (Fig. \ref{fig:combined}). The robot's objective is to deliver a package to the destination, whose reward function balances traveling through difficult terrain, like mud, and reducing the number of actions it takes. To convey its reward function, the robot first provides a human with the demonstration in Fig. \ref{fig:combined}a. Because the robot takes a two-action detour to avoid the mud instead of going through it, the human may infer that the robot associates mud with a negative reward. 

The robot considers what to demonstrate next to convey more information regarding its reward function. Importantly, it knows that the human likely considers mud as costly from the first demonstration, but does not know \emph{how} costly. For instance, the human may counterfactually believe that the robot would take a four-action detour when faced with two mud patches (Fig. \ref{fig:combined}b). However, the robot knows that its ratio of mud to action reward is -3 to -1 and that consequently, it would simply go through the mud in Fig. \ref{fig:combined}b to maximize its reward. Seeing how its direct path meaningfully differs from the human's likely detouring counterfactual (i.e. an alternative, potentially suboptimal behavior), the robot considers this to be an informative next demonstration to provide that targets the human's ZPD -- providing a meaningful yet incremental update to the human belief through an additional unit of information that upper-bounds the cost of mud. 

The robot then follows the two demonstrations with a diagnostic test that simultaneously challenges the human to apply their learned knowledge and reveals whether the robot's current model of the human's beliefs needs to be corrected (Fig. \ref{fig:combined}c). If the human answers incorrectly, the robot may provide feedback, a remedial demonstration, then a sequence of remedial tests and feedback until the human demonstrates concept mastery, inspired by the testing effect (Fig. \ref{fig:combined}e-h). Importantly, the robot continues to update its model of the human's beliefs according to the test answer and throughout the remedial interactions to consider the right counterfactuals when estimating the informativeness of future demonstrations. 





The above interaction demonstrates the importance of maintaining a calibrated model of the human's beliefs through closed-loop testing, which can help select demonstrations that are within the human's ZPD. Our contributions are thus as follows: First, a closed-loop teaching framework that provides demonstrations, tests, and feedback based on insights from the education literature. Second, a particle filter model of human beliefs that supports iterative updates and a calibrated prediction of the counterfactuals likely considered by the human for each demonstration that could be provided. Third, a user study that finds that our framework reduces the regret of human test responses by 43\% over a baseline.


\section{Related Work}

\textbf{Explainable RL:} The field of explainable reinforcement learning (RL) focuses on assisting humans in understanding the decision making of RL agents. Recent surveys ~\cite{milani_fang_2023,puiutta2020explainable,wells2021explainable} highlight a variety of approaches, such as approximating a black box RL policy via an interpretable model (e.g. a decision tree \cite{silva2020optimization}), using saliency maps to highlight features of a state used for decision making \cite{greydanus2018visualizing}, 
and identification of critical training points (e.g. for estimating Q-values \cite{gottesman2020interpretable}). In this work, we focus on a complementary direction to those outlined above -- conveying an understanding of an agent's overall behavior through representative examples.


\textbf{Policy Summarization:} Policy summarization aims to provide a global understanding of a policy to a user through example state-action pairs \cite{Amir2019}, which can aid in transparency. One approach relies on heuristics such as entropy or differences in Q-values to select states and actions to show \cite{huang2018establishing,amir2018highlights}. We instead build on the second approach based on machine teaching \cite{zhu2015machine}, which we highlight below. 


Lee et al. model human learning from agent demonstrations as resembling inverse reinforcement learning (IRL), and leverage human teaching techniques such as scaffolding~\shortcite{lee2021machine} and principles from cognitive science such as counterfactual reasoning~\shortcite{lee2022reasoning} to provide demonstrations that incrementally provide information on the agent's underlying reward function. However, these methods model the human learner as using exact IRL \cite{ng2000algorithms}, which is unable to gracefully handle conflicting information (e.g. knowledge assumed to be learned but failed to be demonstrated during testing). Furthermore, they utilize tests for assessment only after having provided demonstrations. We build on this line of work by proposing a Bayesian model of human beliefs in the form of a particle filter and also utilizing intermittent testing to simultaneously maintain an up-to-date model of human beliefs and provide targeted instruction.

Huang et al.~\shortcite{huang2019enabling} also use Bayesian IRL \cite{ramachandran2007bayesian} to model human learning from agent demonstrations, but only update the relative probabilities of a static set of reward beliefs with each additional demonstration. We instead allow for resampling \cite{li2013adapting} of the beliefs within our particle filter to more efficiently approximate the posterior distribution of human beliefs. Finally, Qian and Unhelkar~\shortcite{qian2022evaluating} also explore interactive policy summarization where they allow humans to request specific demonstrations from an agent. They find that a hybrid strategy of AI-selected and human-selected demonstrations yields the best objective and subjective results; our proposed approach could supply the former demonstrations in their framework.

\section{Technical Background}

This section provides the background for selecting informative demonstrations for a (human) learner using IRL-like reasoning to infer a reward function underlying demonstrations.

\textbf{Markov decision process:} The agent models its world as an instance (indexed by $i$) of a Markov decision process, $MDP_i$, comprised of sets of states $\mathcal{S}_i$ and actions $\mathcal{A}$, a transition function $T_i$, reward function $R$, discount factor $\gamma$, and initial state distribution $S^0_i$. We refer to a group of related MDP instances as a \textit{domain} (described below) and \(\mathcal{S}: \bigcup_i \mathcal{S}_i\) is the union over all of their states. An optimal trajectory $\xi^*$ is a sequence of $(s_i, a, s_i')$ tuples that follow the agent's optimal policy $\pi_i^*$. In line with prior work \cite{abbeel2004apprenticeship}, reward $R$ is represented as a weighted linear combination of reward features $\phi$: $R = \mathbf{w^*}^\top \phi(s, a, s')$.
Finally, we assume the human is aware of the full MDP apart from weights $\mathbf{w^*}$. 

A domain is a group of MDPs that share $R, \mathcal{A},$ and $\gamma$ but differ in $T_i, \mathcal{S}_i,$ and $S^0_i$. For example, all MDPs in the delivery domain share the same $R$ even though they may contain different mud patches (Figs. \ref{fig:combined}a and \ref{fig:combined}b). Thus through IRL, all demonstrations within a domain will support inference over a common $\mathbf{w}^*$. We simplify the notation such that $\pi^*$ refers to any optimal policy within a domain, and $\xi^*$ refers to a demonstration (dropping the corresponding MDP). 

\textbf{Machine teaching for policies:} Our objective in selecting an informative curriculum of demonstrations for conveying $\pi^*$ is captured by the machine teaching framework for policies \cite{lage2019exploring}. We aim to select a set of demonstrations that helps a human, who is assumed to use IRL-like reasoning \cite{jara2019theory}, approximate $\mathbf{w}^*$ and then perhaps use planning \cite{shteingart2014reinforcement} to recover $\pi^*$. Thus, the objective reduces to selecting demonstrations that are informative at conveying $\mathbf{w}^*$, which can be measured using behavior equivalence classes.

\textbf{Behavior equivalence class:} The \textit{behavior equivalence class} (BEC) of a demonstration is the set of reward functions under which the demonstration is still optimal.

For a reward function that is a weighted linear combination of features, the BEC of a demonstration $\xi^*$ of $\pi^*$ is defined as the half-space \cite{lee2022reasoning} formed by the exact IRL equation \cite{ng2000algorithms}
\begin{equation}
\textrm{BEC}(\xi^*|\pi^*, \pi_{\mathbf{w}}) := \mathbf{w^*}^{\top}\left(\mu_{\pi^*}^{s}-\mu_{\pi_{\mathbf{w}}}^{s}\right) \geq 0, s = \xi^*(0),
    \label{eq:BEC_demo}
\end{equation} 
where $\mu^{s}_\pi= \mathbb{E}\left[\sum_{t=0}^{\infty} \gamma^{t} \phi\left(s_{t}\right) \mid \pi, s_0 = s \right]$ is the vector of reward feature counts accrued from starting in $s$ and following $\pi$ after ($\pi_{\mathbf{w}}$ is the optimal policy under reward weight $\mathbf{w}$) and $\xi^*(0)$ is the first state of $\xi^*$.
Any demonstration can be converted into a constraint on $\mathbf{w^*}$ using Eq. \ref{eq:BEC_demo} and a candidate belief $\mathbf{w}$. Importantly, each constraint can be considered a \textit{knowledge component} (KC) \cite{koedinger2012knowledge} which captures a characteristic of the reward function (e.g. a tradeoff between the underlying reward feature weights).

Consider again the delivery domain, which has binary reward features $\mathbf{\phi} =$ [\emph{traversed mud}, \emph{battery recharged}, \emph{action taken}], $\mathbf{w^*} \propto [-3, 3.5, -1]$\footnote{In practice, we require $||\mathbf{w^*}||_2 = 1$ to bypass both the scale invariance of IRL and the degenerate all-zero reward function.}. We assume that the human begins with a prior that the weight of the ``action taken" feature is negative (e.g. a bias toward the shortest path, Fig. \ref{fig:demo_pf}a). The demonstration in Fig. \ref{fig:demo_pf}b yields the constraint (or KC) in Fig. \ref{fig:demo_pf}c, which indicates that $w^*_0 \leq 2w^*_2$ (i.e. mud is at least twice as costly as an action), since two actions were taken to detour around the mud rather than counterfactually going through it (the optimal trajectory for a candidate belief that considers mud to be slightly negative, neutral, or slightly positive).  


\section{Methods}

The example of the delivery robot in Section \ref{sec:introduction} highlights the importance of maintaining an up-to-date model of human beliefs and likely counterfactuals when selecting a demonstration. In this section, we propose a particle filter-based model of human beliefs amenable to iterative Bayesian updates and sampling for counterfactual reasoning, where each particle represents a potential human belief regarding the agent's reward function. We then leverage this model in a closed-loop teaching framework that leverages insights from the education literature to select demonstrations that target gaps identified through testing.

\subsection{Particle Filter Human Model}
\label{sec:pf}

\begin{algorithm}
\footnotesize
\caption{Particle Filter for Modeling Human Beliefs}
\begin{algorithmic}[1]
\State Initialize particles $x_{0}^{(i)} \sim p(x_{0})$ for $i = 1,\dots,N$
\For{$t = 1,\dots,T$}  
    \State{// Update filter given new demonstration or test at $t$}
    \For{$i = 1,\dots,N$}
        \State Compute weight $\check{w}_{t}^{(i)} = \check{w}_{t-1}^{(i)} \cdot p(x_{t}^{(i)}| y_{t})$ 
    \EndFor
    \If{$\sum_{j=1}^{N} \check{w}_{t}^{(j)} < \check{w}_{threshold}$}
    \State Perform a particle filter reset  \label{alg:reset}
    \EndIf
    \State Normalize weights $\tilde{w}_{t}^{(i)} = \frac{\check{w}_{t}^{(i)}}{\sum_{j=1}^{N} \check{w}_{t}^{(j)}}$ \label{alg:weight_normalization}

 \State Compute effective sample size $n_\text{eff} = \frac{1}{\sum_{i=1}^N (\tilde{w}_t^{(i)})^2}$
    \If{$n_\text{eff} < N_\text{threshold}$}
    \State \parbox[t]{\dimexpr\linewidth-\algorithmicindent}{%
          Resample $x_{t}^{(i)}$ with probabilities $\tilde{w}_{t}^{(i)}$ using
          \\  KLD resampling \label{alg:resampling}
        }
    \EndIf
\EndFor 
\end{algorithmic}
\label{alg:pf}
\end{algorithm}

The particle filter routines detailed in the following paragraphs come together in Alg. \ref{alg:pf}.

\textbf{Updating Particle Positions and Weights:} Assume a set of particles, defined by their positions and associated weights \{$\mathbf{x}_t, \check{\mathbf{w}}_t$\}. Without loss of generality, assume that a demonstration or test response is provided at each time step $t$. Each demonstration generates multiple constraints by comparing the demonstration against possible counterfactuals trajectories and each incorrectly answered test will generate a single constraint by comparing the true test answer against the incorrect answer, both through Eq. \ref{eq:BEC_demo}. Each constraint generated via a demonstration or a test response is a half-space constraint, with one side being \textit{consistent} with the demonstration or test response and the other side being \textit{inconsistent}. 

\begin{figure*}
\centering
\includegraphics[width=0.86\textwidth]{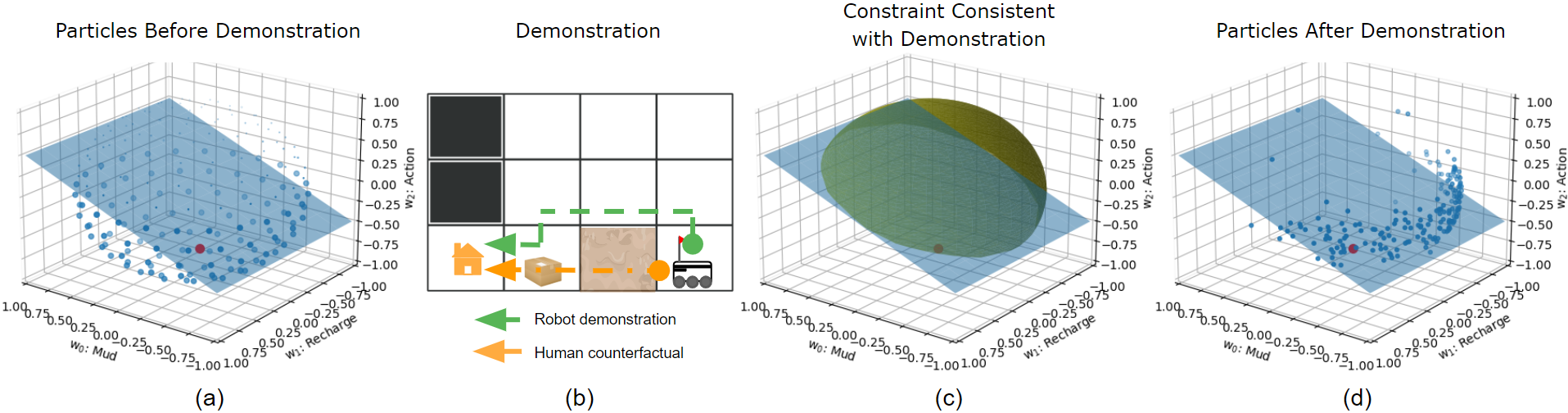} 
\caption{Example sequence on how a demonstration updates a particle filter model of human beliefs. The robot reward function is shown as a red dot, and the constraint consistent with the demonstration is shown in all plots for reference. \textbf{(a)} Particles before demonstration (prior). \textbf{(b)} Demonstration shown to the human, alongside a counterfactual that considers mud to be slightly negative or positive, or neutral. \textbf{(c)} The constraint (Eq. \ref{eq:BEC_demo}) consistent with the demonstration that conveys that mud must be at least twice as costly as an action, visualized with the uniform distribution portion of the custom distribution (Fig. \ref{fig:uniform_vmf}) used to update particle weights.  \textbf{(d)} Particles after demonstration (posterior).}
\label{fig:demo_pf}
\vspace{-4mm}
\end{figure*}

\begin{figure}
\centering
\includegraphics[width=0.44\columnwidth]{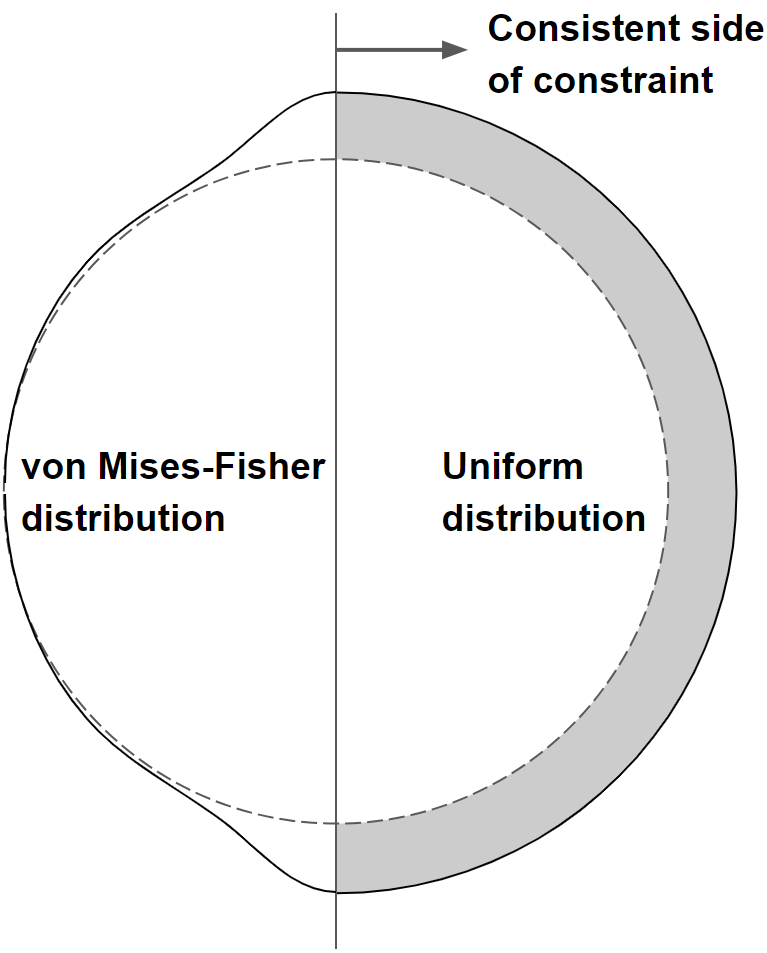} 
\caption{Cross-section of the spherical probability density function used to update particle weights given a constraint generated from a demonstration (Eq. \ref{eq:BEC_demo}).
}
\label{fig:uniform_vmf}
\vspace{-5mm}
\end{figure}

Each constraint $y_t$ can then be translated into a probability distribution $p(x_t | y_t)$ that can be used to update the weights of each particle (Fig. \ref{fig:demo_pf}). We propose a custom probability distribution $p(x_t | y_t)$ that translates each constraint into a combination of a uniform distribution that aligns with the consistent half-space of the constraint and a von Mises-Fisher distribution (a generalization of the Gaussian distribution on a sphere \cite{dhillon2003modeling}) whose mean direction aligns with the inconsistent half-space (Fig. \ref{fig:uniform_vmf}). The uniform distribution asserts that any particle lying on the consistent half-space is equally valid for that demonstration, whereas the Von-Mises Fisher distribution asserts that a particle is exponentially less likely to have generated that demonstration as you move away from the consistent side of the constraint. The resulting probability density function (pdf) of the custom distribution is given in Eq. \ref{eq:custom_dist}, with the normalizing constant $c_1$ that ensures that the pdf sums to 1 (Eq. \ref{eq:custom_dist_c1}), and the scaling constant $c_2$ that matches the probability of the Von-Mises Fisher distribution to that of the uniform distribution at meeting point of the two distributions (Eq. \ref{eq:custom_dist_c2}). Though the custom distribution naturally generalizes to higher dimensions, the particles in our domains with three reward features are constrained to the 2-sphere and the pdf is specified accordingly.

\begin{equation}
f(x, \mu, \kappa) =
    \begin{cases} 
      \frac{1}{2 \pi c_1}, & \mu^\top x \geq 0\\
      \frac{c_2 \kappa e^{\kappa \mu^\top x}}{2 c_1 \pi (e^\kappa - e^{-\kappa})}, &\mu^\top x < 0
   \end{cases} \\
   \label{eq:custom_dist}
   \end{equation}
\begin{equation}
   c_1 = \frac{1}{c_2 \displaystyle \int_0^\pi \int_{\frac{\pi}{2}}^{\frac{3 \pi}{2}} \frac{\kappa e^{\kappa cos(\theta) \cdot sin(\phi)}sin(\phi)}{2 \pi (e^\kappa - e^{-\kappa})} d\theta d\phi + 0.5} 
       \label{eq:custom_dist_c1}
   \end{equation}
\begin{equation}
c_2 = \frac{1}{4 \pi f(y, \mu, \kappa)}, \forall y \, \, s.t. \, \mu^\top y = 0
    \label{eq:custom_dist_c2}
\end{equation}

\textbf{Sampling Human Beliefs:} Given a running particle filter model, we may sample human beliefs in order to do counterfactual reasoning over how the human may interpret each demonstration that could be shown. We first run systematic resampling on a copy of the particles to downselect to a candidate set, favoring those that are higher weighted. We then rely on the 2-approximation algorithm \cite{hochbaum1985best} to greedily select $k$ distributed samples such that the maximum distance from any particle in the candidate set to one of the $k$ samples is minimized. The algorithm iteratively picks the particle with the largest distance to the already selected samples as the next sample; this heuristic ensures that the maximum distance from any particle to any of the selected samples is never worse than twice the optimal. As nearby particles are likely to generate similar counterfactuals, we sample beliefs that are approximately spread out. 

Due to space constraints, please find practical tips on how to resample the particle filter to combat sample degeneracy and impoverishment (line \ref{alg:resampling} of Alg. \ref{alg:pf}), as well as how to reset the particle filter if it receives heavily conflicting information (line \ref{alg:reset} of Alg. \ref{alg:pf}) in the supplementary material.


\subsection{Closed-loop Teaching}

\begin{figure*}[h]
\centering
\includegraphics[width=0.83\textwidth]{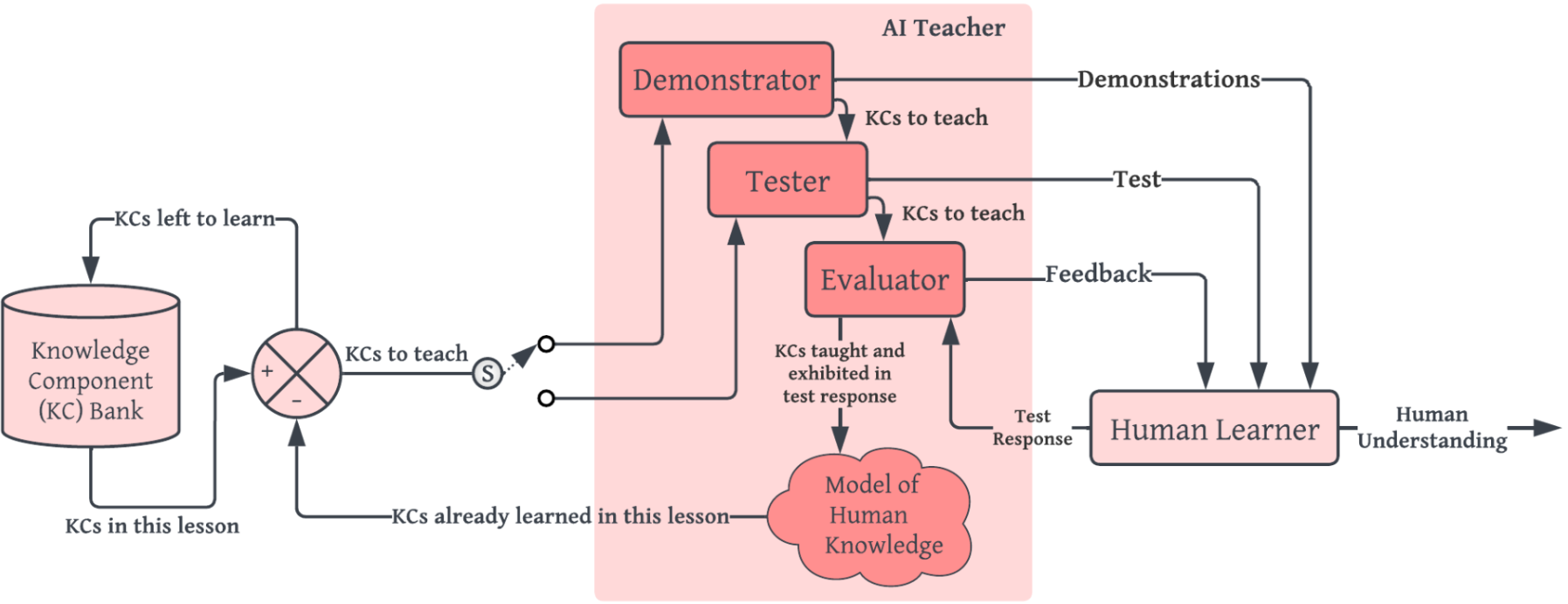} 
\caption{Proposed closed-loop teaching framework. Knowledge components (KCs) are passed to the AI teacher as a lesson. The \texttt{demonstrator} generates demonstrations that convey the KCs, the \texttt{tester} provides test(s), and the \texttt{evaluator} analyzes the test response(s), provides feedback on its correctness, and updates the model of human knowledge. If the human fails to learn a KC through two rounds of demonstrations and tests, the switch (labeled `S') flips such that only tests and feedback are provided until an understanding of the remaining KCs is demonstrated through correct responses.}
\label{fig:closed_loop}
\vspace{-4mm}
\end{figure*}

With a particle filter model of human beliefs amenable to iterative updates via demonstrations and tests, we now formulate a closed-loop teaching framework for conveying an agent's reward function to a human. As we walk through the framework conceptualized in Fig. \ref{fig:closed_loop}, we highlight the principles from the education literature that guide the design. A sample rollout of a teaching sequence is shown in Fig. \ref{fig:combined}, which serves as a visual correspondence to the algorithmic characterization of the framework in the supplementary material.


We first leverage feature and counterfactual scaffolding from Lee et al.~\shortcite{lee2022reasoning} to select KCs that incrementally increase in information across an increasing subset of features (e.g. mud vs action cost, recharging vs action cost, then tradeoffs between all three). This set of KCs guides the machine teaching selection of the \textit{curriculum} of demonstrations that can be used to teach the agent reward function to a human.

We begin the loop by taking a single batch of related KCs that define a \textit{lesson} (e.g. bounds on mud cost) and providing it to the \texttt{demonstrator} (Fig. \ref{fig:closed_loop}) to select demonstrations from the curriculum that convey these KCs. Specifically, we utilize counterfactual reasoning \cite{lee2022reasoning} to select demonstrations that are informative with respect to the counterfactuals likely considered by the human. We simultaneously leverage the educational principles of the \textit{ZPD} \cite{vygotsky1980mind} to provide a sequence of demonstrations that provide information incrementally, i.e. demonstrations that convey one new constraint at a time (such as first providing a lower-bound on the mud cost, then later an upper-bound). 

After the demonstrations have been provided, the \texttt{tester} selects \textit{diagnostic tests} that will verify whether the human has learned the KCs in the lesson. These diagnostic tests optimize for visual dissimilarity from the teaching demonstrations and visual complexity (i.e. increasing distracting visual clutter) \cite{lee2022reasoning} to challenge the learner. 

For each diagnostic test that is answered incorrectly, the \texttt{evaluator} will provide immediate \textit{feedback} to the human on how their answer differed from the correct one, inspired by findings that immediate feedback on errors leads to better learning outcomes \cite{koedinger2013instructional}. And for each diagnostic test that is answered incorrectly, a remedial demonstration that most closely conveys the missed KC with visual simplicity \cite{lee2021machine} will be provided to focus on the concept being taught, along with a remedial test with visual complexity to challenge the learner in demonstrating the missed KC. We note that this missed KC is determined by comparing the human's test answer with the optimal test answer; while it may or may not be the same as one of the KCs originally contained in the lesson, it best addresses the human's current misunderstanding. If the human also gets the remedial test wrong, the switch in Fig. \ref{fig:closed_loop} (labeled `S') flips, and the \texttt{tester} and \texttt{evaluator} will continue to provide only visually dissimilar and complex remedial tests with corresponding feedback (but no additional demonstrations) until the human shows understanding of each iteration's missed KC. This is motivated by the testing effect \cite{roediger2006power}, which supports using tests not only for assessment but also for teaching and increasing learning outcomes. Note that for each demonstration provided or test response received throughout this learning process, we update the particle filter model of the human's beliefs. And we utilize the particle filter model to consider the counterfactuals the human is likely to consider for each potential remedial demonstration or remedial test in order to select the one that will best convey or test the missed KC for the human. Once all of the missed KCs for this lesson have been demonstrated via correct remedial test responses, a fresh batch of KCs (i.e. a new lesson) is pulled from the KC bank and the switch flips upward to provide demonstrations again. 

Alternatively, if all diagnostic tests in this lesson had been correctly answered initially, a fresh batch of KCs would have been pulled from the KC bank to begin the next lesson directly without remedial instruction. 

When all lessons have been taught, the human's subsequent knowledge can be evaluated on a held-out set of tests in which they predict the policy in previously unseen environments.

\section{User Study}

We ran an online user study\footnote{Code for the study and data analysis, as well as the collected data can be found at \href{https://github.com/SUCCESS-MURI/closed_loop_teaching_study}{\texttt{https://tinyurl.com/2s4xty56}.}} exploring whether our proposed closed-loop teaching method improves the transparency of an agent's policy to a human. The study involved participants learning about the agent policy in two domains through a combination of demonstrations, tests, and feedback and predicting the agent's behavior in new test environments. 

\begin{figure}
\centering
\includegraphics[width=0.95\columnwidth]{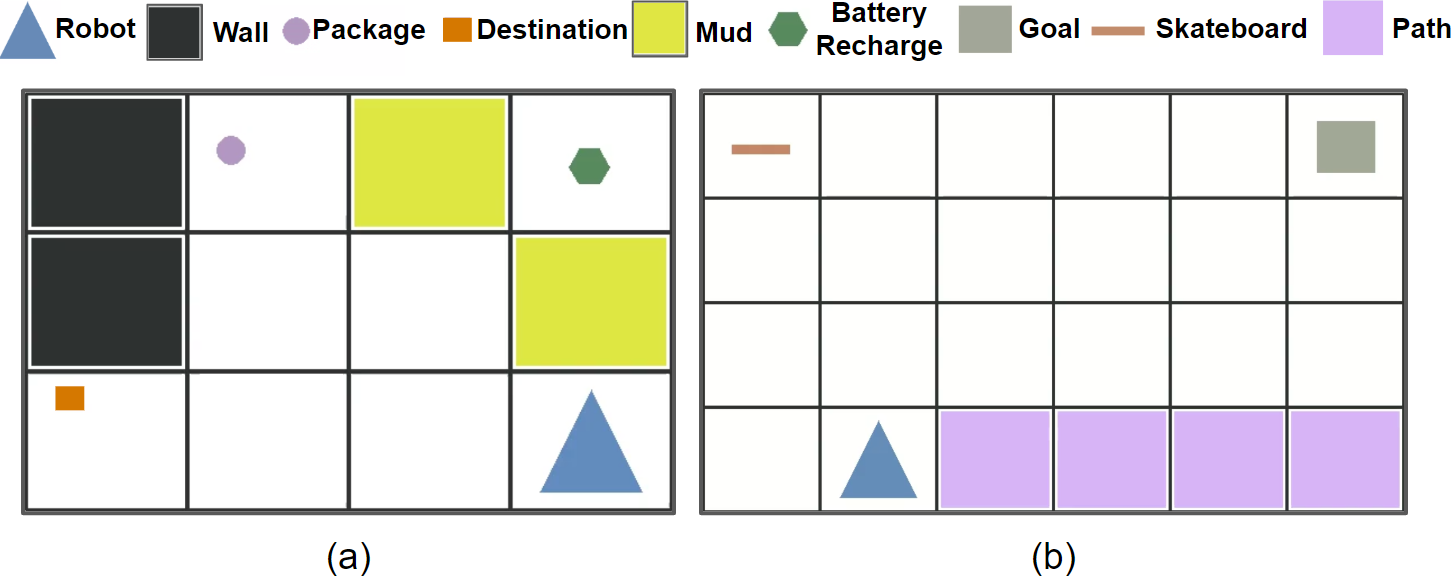} 
\caption{Two domains designed for user study, \textbf{(a)} delivery, \textbf{(b)} skateboard. The semantics of the objects were hidden using arbitrary shapes and colors.}
\label{fig:domains3}
\vspace{-5mm}
\end{figure}

The within-subject variable was \emph{domain}, which consisted of the following two conditions. In the \textit{delivery} domain, the agent is penalized for moving out of mud and rewarded for recharging. In the \textit{skateboard} domain, the agent is rewarded each time it moves with the skateboard (e.g. riding is efficient) or traverses through a designated path (see Fig. \ref{fig:domains3}). Thus each domain consists of two unique reward features and one shared feature that penalizes each action. The \textit{skateboard} domain was designed to be more challenging than the \textit{delivery} domain (confirmed through pilot studies and study results), as the value of the skateboard depends both on the distance to the skateboard and subsequent distance to the goal. 

The between-subjects variable was \emph{feedback loop} with the following three conditions. \textit{Open} feedback loop followed prior work by Lee et al.~\shortcite{lee2022reasoning} in utilizing counterfactual reasoning to select a set of informative demonstrations a priori that incrementally decreased in cumulative BEC area (i.e. a model of human beliefs), one KC at a time. \textit{Partial} feedback loop additionally provided a diagnostic test after each lesson and provided feedback as necessary, while the \textit{full} feedback loop additionally provided a remedial demonstration and remedial tests until the KC in question was correctly applied in a remedial test. For a fair comparison, each condition showed the same median number of demonstrations and tests (11 for delivery and 22 for skateboard).  


The user study consisted of two trials, with each trial comprising a teaching portion and a testing portion in one domain. During teaching, participants were first explicitly informed of the reward features of the domain. Then they inferred the corresponding reward weights by watching demonstrations and perhaps undergoing diagnostic tests, corrective feedback, and further remedial instruction depending on their assigned feedback loop condition. For every interaction, participants indicated whether it improved their understanding of the agent's policy via a Likert scale. At the end of the teaching session, participants were asked to rate their level of focused attention, the usability of their assigned teaching condition, and their understanding of the agent's policy via Likert scales. During testing, participants were tasked with predicting the agent's optimal trajectory in six unseen test environments in random order, which were selected according to prior work \cite{lee2022reasoning} to comprise two low, medium, and high difficulty environments each. We tested the following hypotheses (H1--H4) using the measures (M1--M4) below. 

\noindent \textbf{H1}: (a) The test responses will be best for \textit{full} feedback loop, then \textit{partial}, then \textit{open}. (b) \textit{Delivery} will result in better test responses over \textit{skateboard}.

\noindent \textbf{H2}: (a) Focused attention and perceived usability will be highest for \textit{full} feedback loop, then \textit{partial}, then \textit{open}. (b) \textit{Delivery} will result in higher focused attention and perceived usability over \textit{skateboard}.

\noindent \textbf{H3}: (a) Improvement ratings will be highest for \textit{full} feedback loop, then \textit{partial}, then \textit{open}. (b) \textit{Delivery} will result in higher improvement ratings over \textit{skateboard}.

\noindent \textbf{H4}: (a) Understanding ratings will be highest for \textit{full} feedback loop, then \textit{partial}, then \textit{open}. (b) \textit{Delivery} will result in higher understanding ratings over \textit{skateboard}.

\noindent \textbf{M1. Test response:} The reward of the human's test response, measuring the human's ability to predict the agent's policy.

\noindent \textbf{M2. Focused attention and perceived usability:} We adapted the User Engagement Scale short form \cite{o2018practical} to ask six questions targeting focused attention and perceived usability, each answered with a 5-pt Likert scale.

\noindent \textbf{M3. Improvement:} ``Did this interaction improve your understanding of the game strategy [i.e. agent policy]?'', answered with a 5-pt Likert scale.

\noindent \textbf{M4. Understanding:} ``Do you feel that you now understand the game strategy?'', answered with a 5-point Likert scale.

\section{Results and Discussion}

We collected data from 206 participants using Prolific. Participants were roughly 70\% male, 28\% female, 1\% non-binary, and 1\% preferred not to disclose, and ages varied from 18 to 67 (M = 32.49, SD = 11.15). The recruitment process and study were approved by Carnegie Mellon University's Institutional Review Board. In the \textit{full} feedback loop condition, we removed data from one participant who did not miss any diagnostic tests during teaching (thus did not see any remedial instruction), and one outlier participant whose total number of interactions exceeded 3 standard deviations of the mean number of interactions in this condition (as repeated failures of similar remedial tests suggested lack of attention). This left 68 participants in each between-subjects condition.

\begin{figure}[t]
  \centering
  \includegraphics[width=0.87\columnwidth]{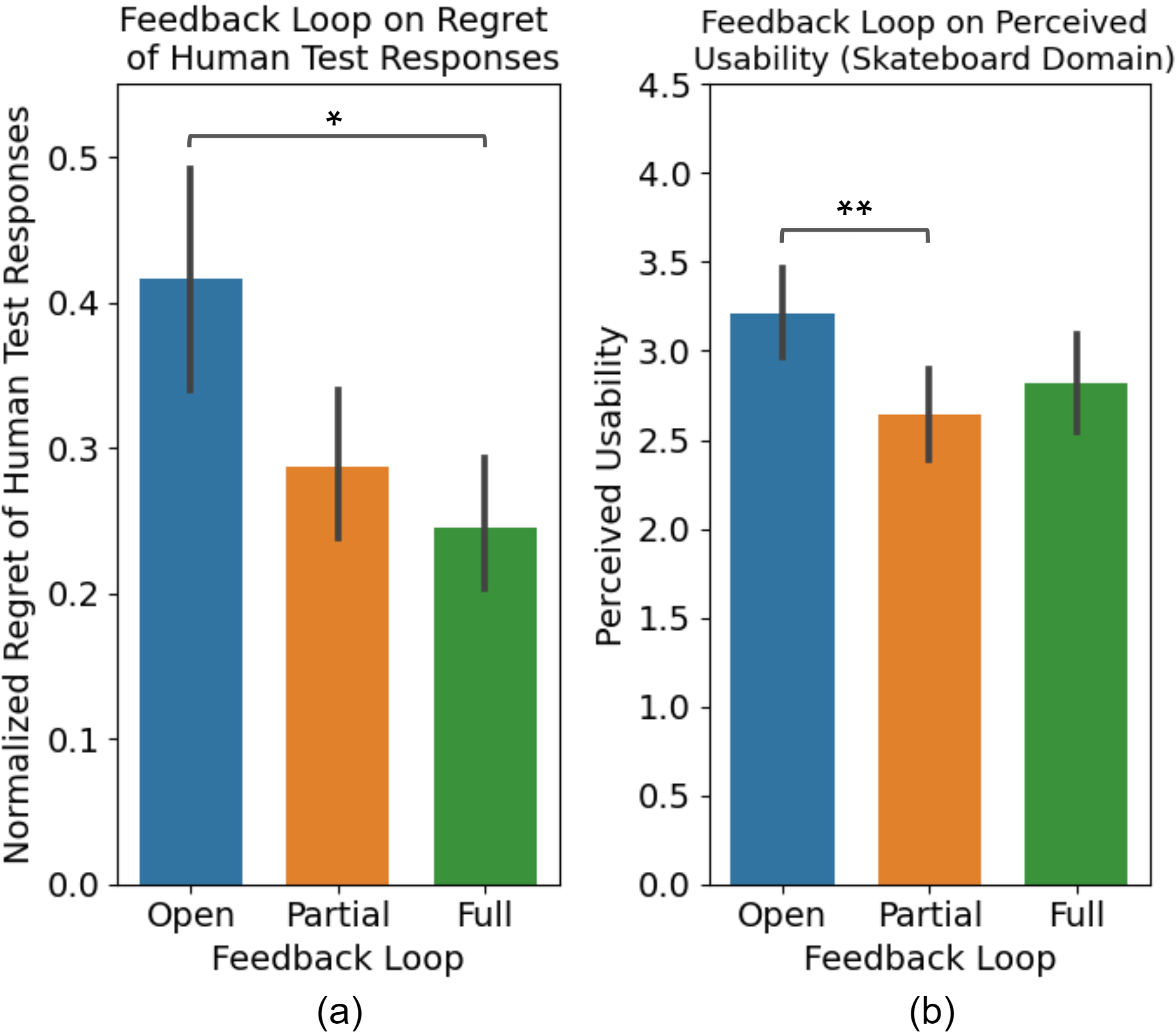}
  \caption{(a) \textit{Full} closed-loop teaching yields lower regret for human tests responses than \textit{open} across domains (lower is better). (b) \textit{Partial} yields lower ratings on perceived usability (higher is better) than \textit{open} in the skateboard domain. Error bars indicate 95\% confidence intervals.}
  \label{fig:H1_H2_interaction}
  \vspace{-5mm}
\end{figure}

\textbf{H1:} We measured the degree of suboptimality of human test responses using regret, i.e. the difference between rewards of human and optimal test responses.  A two-way mixed ANOVA indicated a significant effect of feedback loop on regret ($F(2,201) = 3.65, p = .03$). Tukey analyses revealed that \textit{full} ($M = 0.24$) had 43\% lower regret over \textit{open} ($M = 0.42$, $p = .027$), with \textit{partial} sitting in between with no significant difference to either ($M = 0.29$, Fig. \ref{fig:H1_H2_interaction}a). The ANOVA also indicated a significant effect of domain on regret ($F(1,201) = 50.75, p \leq .001$), where a t-test revealed a significant difference between the regret between \textit{delivery} ($M = 0.18$) and \textit{skateboard} ($M = 0.45$), $t(406) = -5.792, p < .001$. 

The ANOVA also indicated an interaction effect ($F(2, 201) = 3.45, p = .03$) between feedback loop and domain. In the \textit{skateboard} domain, Tukey analyses revealed that \textit{full} ($M = 0.33$) had significantly lower regret over \textit{open} ($M = 0.62$, $p = .014$),


\textit{H1a is partially supported.} Though the regret for \textit{partial} sat in between \textit{full} and \textit{open} as expected (being an intermediary between those two conditions), it was not significantly different from either. However, \textit{full} did indeed significantly outperform \textit{open} even with the same median number of interactions -- highlighting the importance of tailoring the content and interaction type to the human in situ. Finally, the interaction effect reveals that the difference between \textit{full} and \textit{open} on regret is largely driven by the \textit{skateboard} domain, suggesting perhaps that the benefit of the proposed fully closed-loop teaching scheme is greater for more challenging domains. \textit{H1b is supported.}  \textit{Delivery} resulted in a significantly lower regret over \textit{skateboard}, as expected.

\textbf{H2:} A two-way mixed ANOVA did not find a significant effect of feedback loop ($F(2, 201) = 1.56, p = 0.21$), nor domain ($F(1, 201) = 0.38, p = .54$) on focused attention, nor an interaction effect between feedback loop and domain on focused attention ($F(2, 201) = 1.90, p = .15$). A two-way mixed ANOVA found a significant effect of domain on perceived usability ($F(1,201) = 85.77, p < .001$). A t-test revealed a significant difference in the perceived usability ratings of \textit{delivery} ($M = 3.57$) and \textit{skateboard} ($M = 2.89$), $t(406) = 6.562, p < .001$. Finally, a two-way mixed ANOVA also found an interaction effect between feedback loop and domain on perceived usability ($F(2, 201) = 6.17, p = .003$), where Tukey revealed a significant difference between \textit{partial} ($M = 2.64$) and \textit{open} ($M = 3.21)$ for \textit{skateboard} ($p = .006$, Fig. \ref{fig:H1_H2_interaction}b). A main effect of feedback loop on perceived usability was not found ($F(2, 201) = 2.06, p = .13$).

\textit{H2a is not supported.} Though no main effects were found for feedback loop on focused attention or perceived usability, analysis of interaction effects in the \textit{skateboard} domain interestingly reveals that \textit{partial} feedback loop is rated less usable than \textit{open} loop. A number of people in \textit{partial} noted that they wanted more demonstrations to clear up confusion, and we hypothesize that perhaps it can be frustrating to have diagnostic tests highlight gaps in understanding without providing further instruction (as in the case of \textit{full}) or simply providing additional instruction without highlighting gaps in understanding (as in the case of \textit{open}). \textit{H2b is partially supported.} The trend of the domain differences continues with \textit{delivery} yielding significantly higher ratings of perceived usability over \textit{skateboard}, though no difference was found between the domains for focused attention.

\textbf{H3:} As participants gave an \textit{improvement} rating for each demonstration or test, a mean is more descriptive than a median for each participant and for each domain, and we use parametric analyses accordingly. A two-way mixed ANOVA indicated a significant effect of domain on improvement ($F(1,201) = 32.17, p < .001$). A t-test revealed that the teaching in \textit{delivery} ($M = 3.38$) was rated to yield higher improvement than in  \textit{skateboard} ($M = 3.12$), $t(406) = 3.001, p = .003$). The ANOVA did not indicate a significant effect of feedback loop ($F(2, 201) = 1.54, p = .22$) nor a significant interaction effect ($F(2,201) = 1.23, p = .29$) between feedback loop and domain.  


\textit{H3a is not supported.} Feedback loop did not impact ratings of improvement. \textit{H3b is supported.} The ratings suggest 
that participants learned more overall about the \textit{delivery} domain than the \textit{skateboard} domain.

\textbf{H4:} A Kruskal-Wallis H test did not reveal a statistically significant effect of feedback loop on ratings of understanding $(p = .41)$. However, a Wilcoxon signed-rank test showed a statistically significant difference in ratings of understanding between \textit{delivery} and \textit{skateboard} domains $(Z = -6.474, p < .001)$. The mean rating on understanding for \textit{delivery} was 3.90 and the mean rating for \textit{skateboard} was 3.34.

\textit{H4a is not supported}. Feedback loop did not impact ratings of understanding. \textit{H4b is supported.} The ratings support a difference in the difficulty of the two domains.

\section{Conclusion}

Machine teaching provides a principled framework for selecting demonstrations \textit{a priori} that increases the transparency of AI policies to humans; however, individuals may differ in their learning trajectories \textit{in situ}. We thus augment a curriculum of preselected demonstrations with a novel closed-loop teaching framework inspired by key concepts from the education literature to provide tailored instruction. A user study finds that our teaching framework consisting of demonstrations, tests, feedback, and remedial instruction reduces the regret in human test responses by 43\% over a baseline. 


\bibliographystyle{named}
\bibliography{ijcai24}

\end{document}